\documentclass[aps,prb,twocolumn,superscriptaddress, showspace,showkeys,floatfix]{revtex4}
\usepackage{amsmath}
\usepackage{bm}
\usepackage{amssymb}
\usepackage{graphicx}
\usepackage{epstopdf}
\usepackage{bmpsize}
\usepackage{epsfig}
\usepackage{color}
\usepackage{hyperref}
\usepackage{float}
\usepackage{braket}
\usepackage{lipsum}

\begin{document}
\title{Limit Cycle Phase and Goldstone Mode in Driven Dissipative Systems}

\author{H. Alaeian}
\affiliation{5th Institute of Physics, University of Stuttgart, Pfaffenwaldring 57, 70569 Stuttgart, Germany}
\affiliation{Center for Integrated Quantum Science and Technology (IQST), University of Stuttgart, Pfaffenwaldring 57, D-70550 Stuttgart, Germany}

\author{G. Giedke}
\affiliation{Donostia International Physics Center, Paseo Manuel de Lardizabal, 4
E-20018 Donostia-San Sebastian, Spain}
\affiliation{Ikerbasque Foundation for Science, Maria Diaz de Haro 3, E-48013 Bilbao, Spain}

\author{I. Carusotto}
\affiliation{INO-CNR BEC Center and Department of Physics, University of Trento, I-38123 Povo, Italy}

\author{R. L\"ow}
\affiliation{5th Institute of Physics, University of Stuttgart, Pfaffenwaldring 57, 70569 Stuttgart, Germany}
\affiliation{Center for Integrated Quantum Science and Technology (IQST), University of Stuttgart, Pfaffenwaldring 57, D-70550 Stuttgart, Germany}

\author{T. Pfau}
\affiliation{5th Institute of Physics, University of Stuttgart, Pfaffenwaldring 57, 70569 Stuttgart, Germany}
\affiliation{Center for Integrated Quantum Science and Technology (IQST), University of Stuttgart, Pfaffenwaldring 57, D-70550 Stuttgart, Germany}

\date{\today}

\begin{abstract}
In this article, we theoretically investigate the first- and second-order quantum dissipative phase transitions of a three-mode cavity with a Hubbard interaction. In both types, there is a MF limit-cycle phase where the local U(1)-symmetry and the time-translational symmetry (TTS) of the Liouvillian super-operator are spontaneously broken (SSB). This SSB manifests itself through the appearance of an unconditionally and fully squeezed state at the cavity output, connected to the well-known Goldstone mode. By employing the Wigner function formalism hence, properly including the quantum noise, we show that away from the thermodynamic limit and within the quantum regime, fluctuations notably limit the coherence time of the Goldstone mode due to the phase diffusion. 
Our theoretical predictions suggest that interacting multimode photonic systems are rich, versatile testbeds for investigating the crossovers between the mean-field picture and quantum phase transitions. A problem that can be investigated in various platforms including superconducting circuits, semiconductor microcavities, atomic Rydberg polaritons, and cuprite excitons.
\end{abstract}

\keywords{Dissipative phase transitions, Goldstone mode, Limit cycle, Strongly interacting photons}

\maketitle

\section{Introduction}
For many decades, quantum phase transitions (QPT) have been the subject of intense studies in several areas of physics~\cite{Sachdev2011}. In a closed system with unitary dynamics, the hallmark of an equilibrium QPT is the  non-analytic behavior of an observable upon changing a physical parameter~\cite{Vojta2000,Greiner2002,Brown2017}. 
In recent years, a new frontier has emerged in many-body physics, investigating non-equilibrium phase transitions. In that regard and as a suitable testbed, driven-dissipative quantum systems and their phase transitions have been the subject of many studies. Observation of exciton-polariton BEC in semiconductors~\cite{Deng2010,Carusotto2013} and cuprites~\cite{Bao2019} and their superfluidity~\cite{Amo2009,Lerario2017}, probing the first-order phase transitions, dynamical hysteresis, and Kibble-Zurek quench mechanism in microcavities~\cite{Rodriguez2017,Fink2018}, and demonstration of dynamical bifurcation and optical bistability in circuit QED ~\cite{Siddiqi2005,Yin2012,Liu2017,Fitzpatrick2017,Elliott2018, Andersen2020} are a few examples of rapidly growing body of experimental explorations of such physics in different platforms.

In parallel, some general aspects of non-equilibrium QPT have been investigated theoretically~\cite{Diehl2010,Torre2013}, and particularly e.g. in coupled spins~\cite{Kessler2012}, interacting bosonic systems~\cite{Casteels2016,Boite2017,Casteels2017, Verstraelen2020}, and semiconductor microcavities~\cite{Carusotto2005}. Due to their coupling to a bath, driven-dissipative dynamics are not given by a Hermitian Hamiltonian but with a superoperator, whose gapped spectrum signifies a QPT~\cite{Drummond1980,Drummond1981,Carmichael2015}. In spite of all progress, due to a notably larger parameter space compared to the closed systems, dissipative phase transitions (DPT) necessitate further investigations. A natural question e.g., could be about the crossover between the DPT and the phase transition in the thermodynamic limit (TD). Although due to their constant interaction with the environment, open systems are inherently far from the thermodynamic equilibrium however, still there could be some parameter ranges where the system asymptotically approaches the mean-field (MF) limit, where quantum correlations and fluctuations can be ignored. 

To be more specific, in this paper we focus our studies on a driven-dissipative three-mode bosonic system subject to a Kerr-type intra- and intermodal interactions. To keep our results and discussions general, we do not specify the nature of the bosonic system. But let us remark that such setup could be realized in various platforms, including cavity Rydberg polaritons~\cite{Jia2018,Schine2019,Clark2020}, excitons in 2D materials and semiconductors~\cite{Togan2018,Tan2020}, microwave photons in superconducting circuits~\cite{Materise2018}, or interacting photons in optical cavities~\cite{Klaers2010}.

Starting from the MF description we first explore the phase transitions of the system as a function its parameters, i.e. pump, detuning, interaction strength, and bare-cavity mode spacing. We show that depending on the bare cavity features, the phase transition can be either continuous (2$^{nd}$-order phase transition) or abrupt (1$^{st}$-order phase transition) corresponding to an optical multi-stability, as studied for planar microcavities~\cite{Wouters2007B}. In both cases, the phase transition manifests itself by a non-zero amplitude of the unpumped modes and is related to the dissipative gap closure of the Liouvillian. We show that within this range and up to the MF level, there is an unconditionally squeezed mode at the output, attributed to the spontaneous breaking of the local U(1)- and time-translational symmetry (TTS). While at TD limit, the diverging quadrature of this state is related to the well-known, freely propagating Goldstone mode~\cite{Wouters2007,Leonard2017,Guo2019}, employing the Wigner phase-space representation we show that within the quantum limit this mode becomes susceptible to fluctuations and becomes short-lived. Since employing the Wigner formalism allows us to properly include the quantum noise, we have been able to explore the phase diagram more accurately and beyond MF description. That also helps to delineate the validity range of MF when it comes to the study of QPT. In spite of its simplicity, the investigated system reveals important dynamics of driven-disspative bosonic gases and could be a quintessential model for further exploration of SSB in open many-body systems.

The paper is organized as follows; in Section~\ref{sec:problem} we present the general problem, its MF description in the form of a generalized Gross-Pitaevskii equation (GPE) and the low-energy excitation spectrum determined via Bogoliubov treatment. We also summarize the stochastic formulation of the problem based on the truncated Wigner phase-space method. 
In Section~\ref{sec:results} we present the numerical results of the three-mode cavity where various phase transitions are investigated and discussed. Finally, the last section summarizes the main results of the paper and sets the stage for future directions that can be explored in such systems.
\section{Problem Formulation}\label{sec:problem}
Consider a three-mode cavity with the following Hamiltonian describing the interaction dynamics between the modes ($\hat{a}_{1,2,3}$) 
\begin{align}~\label{eq:3-mode Hamiltonian}
    \hat{H} &= \sum_{n=1}^3 \left(\omega_m \hat{a}_m^\dagger \hat{a}_m + \frac{V_0}{2} \sum_{m}^3 \hat{a}_m^\dagger \hat{a}_n^\dagger \hat{a}_m \hat{a}_n\right)\\
   &+ V_0 (\hat{a}_2^{\dagger ^2} a_1 a_3 + \hat{a}_1^\dagger \hat{a}_3^\dagger \hat{a}_2^2),
\end{align}
where $\omega_m$ is the frequency of the $m^{th}$-mode of the bare cavity and $V_0$ is the interaction strength.

A coherent drive at frequency $\omega_L$ excites the $p^{th}$-mode of the cavity at the rate of $\Omega_0$ as
\begin{equation}~\label{eq:coherent drive}
    \hat{H}_D = \Omega_0(\hat{a}_p e^{+i\omega_L t} + \hat{a}_p^\dagger e^{-i\omega_L t}).
\end{equation}
Assuming a Markovian single-photon loss for the mode-bath coupling, the following Lindblad master equation describes the evolution of the reduced cavity density matrix $\hat{\rho}$ as 
\begin{equation}~\label{eq:master equation}
    \frac{d\hat{\rho}}{dt} = -i\left[\hat{H} , \hat{\rho}\right] + \sum_m \gamma_m \left(2\hat{a}_m\hat{\rho}\hat{a}_m^\dagger - \{\hat{a}_m^\dagger \hat{a}_m , \hat{\rho}\}\right),
\end{equation}
where $\hat{H} = \hat{H}_{ph} + \hat{H}_D$ on the RHS describes the unitary dynamics of the system and the second term captures the quantum jumps and losses of the $m^{th}$-cavity field at rate $\gamma_m$.

Equivalently, we can derive the equations of motion for $\hat{a}_m$ operators and describe the dynamics via Heisenberg-Langevin equations as~\cite{Gardiner2004}
\begin{multline}~\label{eq:Heisenberg-Langevin}
    \dot{\hat{a}}_m = -i\left(\Delta_m -i\gamma_m \right)\hat{a}_m - iV_0 \sum_{nkl}\eta^{mn}_{kl} \hat{a}_n^\dagger \hat{a}_k \hat{a}_l - i\Omega_0 \delta_{mp} + \\
    \sqrt{2\gamma_m} \hat{\xi}_m(t),
\end{multline}
where in the above equation $\Delta_m = \omega_m - \omega_L$ is the frequency of the $m^{th}$-mode in the laser frame, $\eta_{kl}^{mn}$ is the mode-specific prefactor arising from different commutation relations, and $\{\hat{\xi}_m(t)\}$ describe stationary Wiener stochastic processes with zero means and correlations as
\begin{align}~\label{eq:white-noise correlation}
    \braket{\hat{\xi}_m^\dagger(t+\tau) \hat{\xi}_n(t)} = n_{th} \delta(\tau) ~\delta_{mn}, \\ \nonumber
    \braket{\hat{\xi}_m(t+\tau) \hat{\xi}_n^\dagger (t)} = (1 + n_{th})\delta(\tau) ~\delta_{mn},
\end{align}
$n_{th}$ in the above equations is the number of thermal photons at temperature $T$.

For numerical calculations, the dimension of the relevant (few-photon) Hilbert space grows rapidly with increasing number of modes and particle number. Hence, the direct solution of the density matrix in Eq.~(\ref{eq:master equation}) is only possible for a small number of modes and at a low pumping rate $\Omega_0$. For the quantum Langevin equations in Eq.~(\ref{eq:Heisenberg-Langevin}), the two-body interaction generates an infinite hierarchy of the operator moments, making them intractable as well. 

The most straight-forward approach is a classical MF treatment where the correlations are approximated with the multiplication of the expectation values i.e., $\braket{\hat{a}_m \hat{a}_n} \approx \braket{\alpha_m} \braket{\alpha_n}$. These substitutions simplify the equations of motion of the operators' MFs in Eq.~(\ref{eq:Heisenberg-Langevin}) to a set of coupled non-linear equations as 
\begin{multline}~\label{eq:mean-field equations}
    i\dot{\alpha}_m = \left(\Delta_m -i\gamma_m \right)\alpha_m + V_0 \sum_{nkl}\eta^{mn}_{kl} \alpha_n^* \alpha_k \alpha_l + \Omega_0 \delta_{mp}. 
\end{multline}
In the steady state, the mean values are determined as $\dot{\alpha}_m=0$, which is an exact description for the operators' 1$^{st}$-moments. 
In this work, we used the Jacobian matrix to check the dynamical stability of all steady-states. Equation~(\ref{eq:mean-field equations}) is a Gross-Pitaevskii type equation with added drive and dissipation terms.

Although the MF provides a good starting point, information about quantum correlations is lost. To improve this, we replace $\hat{a}_m = \alpha_m + \hat{b}_m$ and linearize Eq.~(\ref{eq:Heisenberg-Langevin}) around MF determined from the steady state of Eq.~(\ref{eq:mean-field equations}). Defining $\hat{B} = \left[\hat{b}\right]$ as fluctuation-operator vector (with $2N$ components), its time evolution is determined as
\begin{equation}~\label{eq:linearized fluctuations EoM}
    \frac{d\hat{B}}{dt} = M\hat{B} + D^{1/2} \hat{\Xi} ,
\end{equation}
where $M$ is the Bogoliubov matrix at the MF $\alpha_m$, $D=\mathrm{diag}(2\gamma_m)$, and $\hat{\Xi}$ is the noise operator vector of the Wiener processes in Eq.~(\ref{eq:Heisenberg-Langevin}). 
As shown in Appendix~\ref{app:Covariance MF-Bog}, from $\hat{B}$ one can directly determine the covariance matrix, $\mathrm{C}_B(\omega)$ whose entries are the stationary two-time correlations of the (zero-mean) operators $\hat{B}_i,\hat{B}_j$
\begin{equation}~\label{eq:spectral response}
    \Gamma_{ij}(\omega) = \mathcal{F} \braket{\lim_{t\to\infty}\hat{B}_i(t+\tau) \hat{B}_j(t)} = \braket{\Tilde{\hat{B}}_i(\omega) \Tilde{\hat{B}}_j(-\omega)},
\end{equation}
where $\mathcal{F}$ represents the Fourier transform of the correlation w.r.t to the delay $\tau$ and $\Tilde{\hat{B}}_i(\omega)$ is the Fourier transform of $\hat{B}_i(t)$. 

Within the Born-Markov approximation, if the 2$^{nd}$-order dynamics is contractive and, in the vicinity of the steady state  it dominates over the higher-order terms, then most of the important correlations can be obtained from the linearized Bogoliubov treatment as in Eq.~(\ref{eq:linearized fluctuations EoM}). This is a self-consistent criterion with $M$ being a negative-definite matrix and is typically satisfied at large particle numbers and weak interactions, as for the TD limit, where MF treatment is well-justified.

To examine the validity of the MF and linearization in the quantum limit of small number of particles, we further employ the Wigner function (WF) representation to express the system dynamics in terms of the analytic quasi-probability distribution $W(\vec{\alpha};t)$~\cite{Wiseman2011,Gardiner2004,Berg2009}. Using Itô calculus, the truncated dynamics of $W$ can be further mapped to a set of stochastic differential equations (SDE)s for $\alpha_m^\pm$ with the following general form (more details can be found in Appendix~\ref{app:Wigner func.})
\begin{equation}~\label{eq:SDE}
    d\alpha_m = A_m dt + \sum_{m'} D_{m,m'} ~ dN_m,
\end{equation}
where $dN_m$ is a complex Wiener process describing a Gaussian white noise.

For any operator $\hat{\mathcal{O}}$, the expectation value of its symmetrically-ordered form, i.e. the equally weighted average of all possible orderings of the $\hat{\mathcal{O}}$ and $\hat{\mathcal{O}}^\dag$, can be obtained as
\begin{equation}\label{expectationvalue-SDE}
    \braket{\hat{\mathcal{O}}}_{sym} = \braket{\braket{\mathcal{O}}},
\end{equation}
where $\braket{\braket{.}}$ stands for the ensemble average over stochastic trajectories.

Before leaving this section, we would like to emphasize that the beyond-MF corrections of GPE in Eq.~(\ref{eq:mean-field equations}) need the effect of the $2^{nd}$ and the $3^{rd}$ normally- and anomalously-ordered correlations. These terms contribute to the MF as \emph{state-dependent} noises. In the truncated Wigner method, there are additional drift terms as well as Langevin forces to  capture those aforementioned quantum-field corrections, partially. While the full dynamics of $W(\vec{\alpha};t)$ in Eq.~(\ref{eq:Fokker-Planck}) is equivalent to the master equation in Eq.~(\ref{eq:master equation}), the truncated Wigner (TW) is an approximation which can only be applied to initially positive WF and preserves this property. It can be interpreted as the semi-classical version of Langevin equations of Eq.~(\ref{eq:Heisenberg-Langevin}). Thus, the TW and its equivalent SDE in Eq.~(\ref{eq:SDE}) might not be able to reproduce the quantum dynamics, fully. However, it goes beyond the MF-Bogoliubov treatment and can describe the generation of non-Gaussian and non-classical states~\cite{Corney2015}.
\section{Results and Discussion}\label{sec:results}
Throughout this section we assume identical field decay rates for all cavity modes, i.e., $\gamma_m = \gamma_0$ and express all other rates normalized to this value. Similarly, time is expressed in units of $\gamma_0^{-1}$. A coherent drive as in Eq.~(\ref{eq:coherent drive}) excites the second mode, i.e. $\hat{a}_2$ hence, the $1^{st}$ and $3^{rd}$ modes are populated, equally (more discussions can be found in Appendix~\ref{app:Covariance MF-Bog}). Thermal fluctuations due to the bath are assumed to be zero, i.e. $n_{th}=0$. Part of the full quantum mechanical calculations are done with QuTip open-source software~\cite{Johansson2012,Johansson2013}. The numerical convergence in each case has been tested by increasing the number of random initialization (MF), random trajectories (SDE), and the truncation number in Fock states (DM) to have a relative error $\approx O(-5)$ in the particle number.
\begin{figure}[htbp]
\centering
\includegraphics[width=\linewidth]{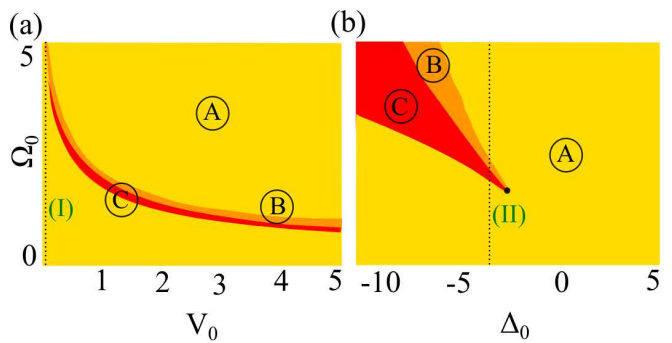}
\caption{~\label{fig:phase transition} MF Dissipative phase diagram of a three-mode harmonic cavity, i.e., $2\omega_2 = \omega_1 + \omega_3$, as a function of (a) the interaction strength $V_0$ and (b) the laser detuning $\Delta_0$. In each panel the yellow (A), orange (B), and red (C) regions correspond to one, two (bi-stability), and three stable (tri-stability) fixed points for the pumped mode, respectively. In (a) the detuning is fixed at $\Delta_0 = -3$ and in (b) the interaction strength has the constant value $V_0 = 1$. The dotted vertical lines [labelled (I) and (II)] at $V_0=0.1$ and $\Delta_0=-3$ indicate the cuts through the phase diagram studied in subsequent figures.}
\end{figure}

In a driven-dissipative system, the interplay between coherent excitation rate and its detuning , incoherent loss , and interaction leads to notable changes in system properties, typically known as dissipative phase transition (DPT). In a multi-mode case as in here, we have an additional parameter $\delta_D = 2\omega_2 - (\omega_1 + \omega_3)$, which is the anharmonicity of the bare cavity. To distinguish between these two cases, we call the cavity \emph{harmonic} if $\delta_D = 0$ and \emph{anharmonic} otherwise. As will be discussed, $\delta_D$ is also an important parameter governing the DPT. Similar phase diagrams and multi-stability phenomena have been studied for exciton-polaritons in planar cavities where $\delta_D$ vanishes~\cite{Wouters2007B}. Moreover, in this case the frequencies of the generated pairs are set by the bare cavity modes and the interaction, self-consistently. 

Figure~\ref{fig:phase transition}(a),(b) shows the phase diagram of a harmonic cavity as a function of the interaction strength $V_0$ and the laser detuning $\Delta_0$, respectively. The phase diagram closely resembles the DPT of a single-mode cavity depicted in Fig.~\ref{fig:pump-only PT}(a),(b) in Appendix~\ref{app:single-mode}. While it is in (A)-phase, i.e. the yellow region, the pumped mode has one stable fixed point. In (B)-phase, i.e. the orange region, there are two distinct values for the pumped mode. Finally in (C)-phase, i.e. the red region which only appears in the multi-mode case, the system is within a tri-stable phase and the pumped mode has three stable MF fixed points.

\begin{figure}[htbp]
\centering
\includegraphics[width=\linewidth]{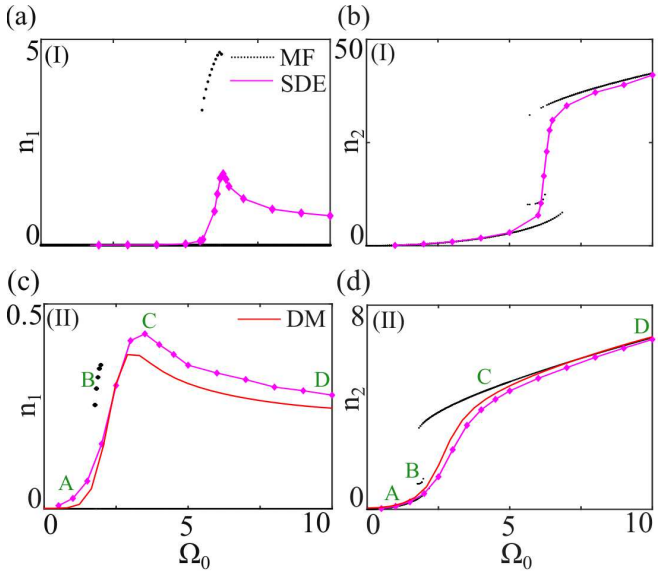}
\caption{~\label{fig:3-mode population-harmonic} Population of the $1^{st}$ ($3^{rd}$) and the $2^{nd}$ mode in a harmonic cavity, i.e., $\delta_D = 0$, as a function of the pumping rate ($\Omega_0$) calculated from MF (black dots) and SDE (purple diamonds). Solid red line in panels (c),(d) show the DM solutions for comparison. $V_0 = 0.1$ in panels (a),(b) and $V_0 = 1$ in (c),(d). In both cases $\Delta_0 = -3$.}
\end{figure}
In Fig.~\ref{fig:3-mode population-harmonic} we plot $\braket{\hat{n}_{1,2}}$ for $V_0 = 0.1,~1$ at $\Delta_0 = -3$ as a function of the pumping rate varied along the dotted lines (I),(II) in Fig.~\ref{fig:phase transition}(a),(b), respectively.
There, the black dots show the MF solutions determined from integrating Eq.~(\ref{eq:mean-field equations}) for many different random initial conditions and for a time long compared to all transient time scales. The purple line with diamonds show the data calculated using the SDE method averaged over 2000 random trajectories, and the solid red line in panel (c),(d) depicts the results of the full density matrix calculations (DM) as a benchmark.
It can be seen that the phase transitions are discontinuous, i.e. a \emph{first-order} PT. Moreover, for all modes the difference between stable MF branches decreases upon increasing the interaction from $V_0 = 0.1$ to $V_0 = 1$ in Fig.~\ref{fig:3-mode population-harmonic}(a,b) and (c,d), respectively. Aside from the finite region around the multi-stability, also it can be seen that the results of MF, SDE, and DM agree quite well (Note a similar tendency for the single-mode case in Fig.~\ref{fig:pump-only} of Appendix~\ref{app:single-mode}). 
For the 1$^{st}$ and 3$^{rd}$ modes on the other hand, both Fig.\ref{fig:3-mode population-harmonic}(a) and (c) indicate that the finite MF tri-stable region (C in the DPT) is the only parameter range where these modes get non-zero population.

\begin{figure}[htbp]
\centering
\includegraphics[width=\linewidth]{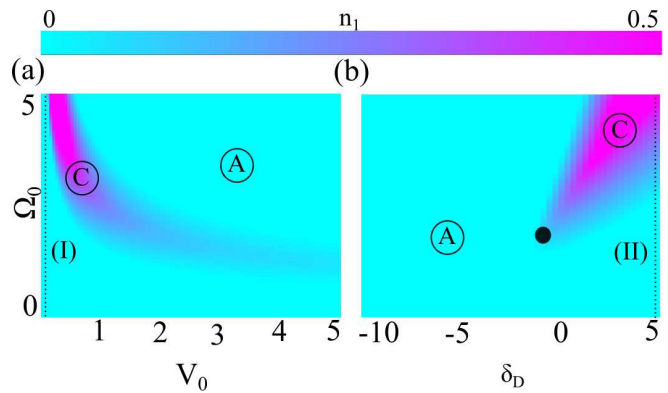}
\caption{~\label{fig:3-mode DPTT anharmonic} Number of photons in the $1^{st},3^{rd}$-modes of a three-mode anharmonic cavity ($2\omega_2 \ne \omega_1 + \omega_3$), as a function of the (a) interaction strength $V_0$ and (b) anharmonicity $\delta_D$, determined from MF. In (a) $\delta_D = 5$ and in (b) $V_0 = 1$ and the laser is always resonantly pumping the $2^{nd}$-mode $\Delta_0 = 0$. (A) , (C) indicate two different phases of zero and non-zero population of the first mode. The dotted vertical lines [labelled (I) and (II)] at $V_0=0.1$ and $\delta_D=5$ indicate the the cuts through the phase diagram studied in subsequent figures.}
\end{figure}

\begin{figure}[htbp]
\centering
\includegraphics[width=\linewidth]{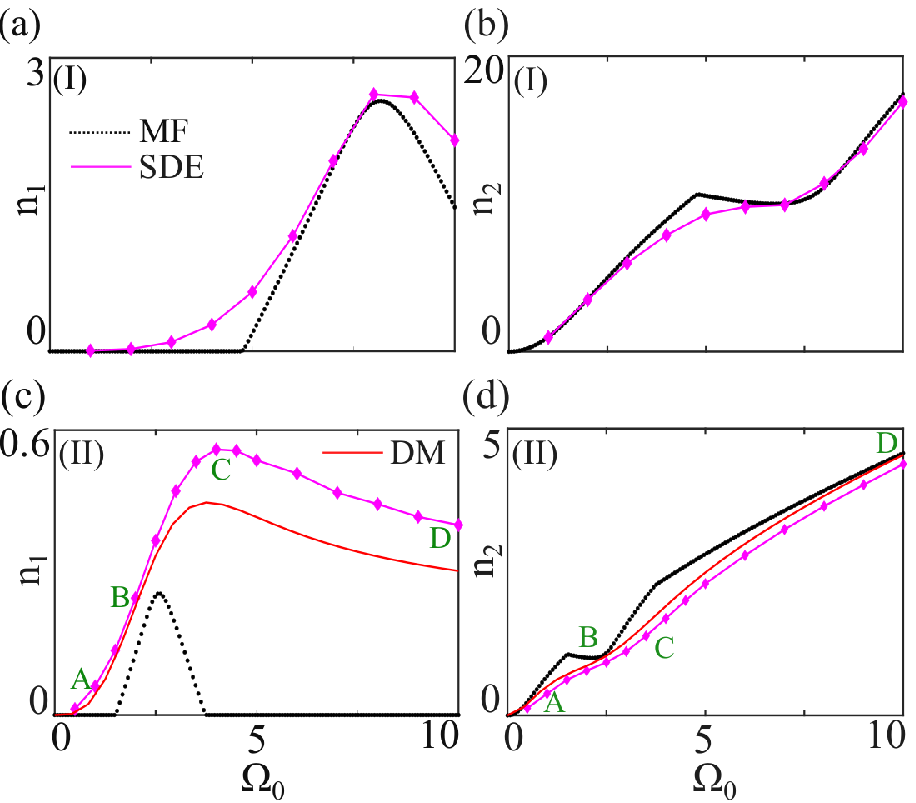}
\caption{~\label{fig:3-mode population-anharmonic} Population of the $1^{st}$ ($3^{rd}$) and the $2^{nd}$ mode, in an anharmonic cavity with $\delta_D = 5$, as a function of the pumping rate ($\Omega_0$) calculated from MF (black dots), SDE (purple diamonds). Solid red line in panels (c),(d) shoe the DM solutions for comparison. $V_0 = 0.1$ in panels (a),(b) and $V_0 = 1$ in (c),(d). In both cases $\Delta_0 = 0$.}
\end{figure}

The situation is completely different in an \emph{anharmonic} cavity where $\delta_D \ne 0$. Figure~\ref{fig:3-mode DPTT anharmonic}(a),(b) shows the average number of photons in unpumped modes $\braket{n_{1,3}}$ as a function of the interaction strength $V_0$, the pumping rate $\Omega_0$, and the anharmonicity parameter $\delta_D$. For better illustrations, in Fig.\ref{fig:3-mode population-anharmonic}(a,b) and (c,d)  we plot the average number of photons in all cavity modes as a function of the pump rate at weak ($V_0 = 0.1$) and strong ($V_0 = 1$) interaction, respectively when the pumping rate is continuously increased along (I) and (II) dotted lines in Fig.~\ref{fig:3-mode DPTT anharmonic}(a),(b). Unlike the harmonic cavity case, here we only have two phases (A),(C), where the transition occurs continuously (but non-analytic), i.e.\emph{second-order} PT, with a unique-valued order parameter in each phase. 

As elaborated in Appendix~\ref{app:Covariance MF-Bog} for the single-mode cavity, the interaction of the pumped mode ($2^{nd}$ mode here) with itself creates energetically symmetric sidebands. In a multi-mode case, the interplay between the intra- and inter-mode interactions leads to the excitation of other modes in both harmonic as well as anharmonic cavities. Similarly for both, MF predicts a threshold and a finite parameter range for non-zero occupations of the the $1^{st}, ~ 3^{rd}$-mode. 
While the lower threshold is set solely by the pumped mode when $V_0 n_2 \ge \gamma_2$, the upper threshold is dependent on the population of the other two modes as well as their relative energies. (The lowest and highest pumping rate is set by the constraints on $\Phi_0 , \Phi_p$, respectively, as detailed in Appendix~\ref{app:Covariance MF-Bog}.)

When quantum fluctuations are included, however, either via SDE or full density matrix calculations (DM), unique, continuous and, non-zero solutions for all three modes are predicted at all pumping rates. In both cavities and for the pumped mode, MF, SDE, and DM results agree quite well in (A)-phase. For the parametrically populated modes however, the SDE and DM results are in good agreement over the whole range but are remarkably different from MF. However, the rising slope of the former analyses always coincide with the transition to the MF (C)-phase. 
\subsection{Spontaneous Symmetry Breaking and Goldstone mode}~\label{sec:SSB}
In the absence of the coherent pump, the Liouvillian super-operator $\mathcal{L}$ of Eq.~(\ref{eq:master equation}) has a continuous global U(1)-symmetry, which is broken by a coherent drive of Eq.(\ref{eq:coherent drive}). However, with the Hamiltonian of Eq.~(\ref{eq:3-mode Hamiltonian}) for the three-mode cavity, $\mathcal{L}$ sill has a local U(1)-symmetry as it remains unchanged under the following transformations for any arbitrary phase $\Theta_0$~\cite{Wouters2007}
\begin{equation}~\label{eq:local U(1)-sym}
    \hat{a}_1 \rightarrow \hat{a}_1 e^{+i\Theta_0} ~ , ~ \hat{a}_3 \rightarrow \hat{a}_3 e^{-i\Theta_0}.
\end{equation}
If the MF amplitudes $\alpha_{1,3} = 0$, then the steady state respects the Liouvillian's symmetry. However, for $\alpha_{1,3} \ne 0$ as occurs within the (C)-phase, the MF solutions are not U(1) symmetric, anymore. Hence, there is a \emph{spontaneous symmetry breaking} (SSB) accompanied by a DPT. However, it is evident that the set of all solutions is invariant under the aforementioned rotations. In other words, within the (C)-phase there is a continuum of MF fixed points. 

\begin{figure}[htbp]
\centering
\includegraphics[width=\linewidth]{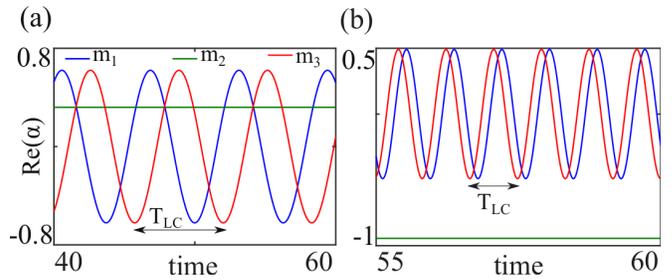}
\caption{~\label{fig:LC} Temporal behavior of the mean fields $\alpha_j(t)$ within the MF (C)-phase in a thee-mode (a) harmonic cavity at $\Omega_0 = 1.85, \Delta_0 = -3$ and (b) anharmonic cavity at $\Omega_0 = 2, \delta_D = 5$ and $V_0 = 1$. In both panels the blue, green, and red lines correspond to the $1^{st},~2^{nd}$ and, the $3^{rd}$-mode, respectively. The time is in units of $\gamma_0^{-1}$ and $T_{LC}$ indicates the limit-cycle period.}
\end{figure}

Figure~\ref{fig:LC}(a),(b) shows the temporal behavior of order parameters $\alpha_m$ within the MF (C)-phase of the harmonic and anharmonic cavities, respectively. As can be seen, while the pumped mode $m_2$ is time-invariant (green line), the parametrically populated modes $m_{1,3}$ (blue and red lines) show self-sustained oscillations with a random relative phase, reflecting the value U(1) phase acquire in the SSB.

In the laser rotated frame, the Liouvillian $\mathcal{L}$ is TTS, which indeed is the symmetry of the solutions within the (A),(B)-phase. Within the (C)-phase, however, the order parameter becomes time-periodic and thus breaks the time-translational symmetry. Therefore, in both of the harmonic and anharmonic cavities, the MF (C)-phase is accompanied by SSB of the local U(1) symmetry and the TTS. This oscillatory behavior, known as \emph{limit-cycle} (LC)-phase, is an apparent distinction of DPT from its equilibrium counterparts~\cite{Qian2012,Chan2015}.
From Fig.~\ref{fig:LC} the LC-period can be determined as $T_{LC} \approx 6.28$ and $T_{LC} \approx 0.83$, corresponding to $\omega_{LC} = 1 , ~ 7.5$ for the harmonic and anharmonic cavities, respectively. Note that these frequencies agree with theoretical predictions of $\Tilde{\Delta}_{1,3}$ in Appendix~\ref{app:Covariance MF-Bog}.

The consequence of SSB of this continuous symmetry can be interpreted in terms of the gapless \emph{Goldstone} mode. 
The eigenvalues $\{\lambda\}$ of the Bogoliubov matrix $M$ in Eq.(\ref{eq:linearized fluctuations EoM}) directly determine the excitation energies around a MF fixed point, with Re($\lambda$) being the excitation linewidth and Im($\lambda$) its frequency. It is straightforward to check that due to the relative-phase freedom of the unpumped modes, $M$ has a kernel along the following direction~\cite{Wouters2007} (more information in Appendix~\ref{app:Covariance MF-Bog})
\begin{equation}~\label{eq:Goldstone mode}
    \ket{G} = [\alpha_1, 0, -\alpha_3, -\alpha_1^*, 0, \alpha_3^*]^T,
\end{equation}
where $T$ means the transpose. 

\begin{figure}[htbp]
\centering
\includegraphics[width=\linewidth]{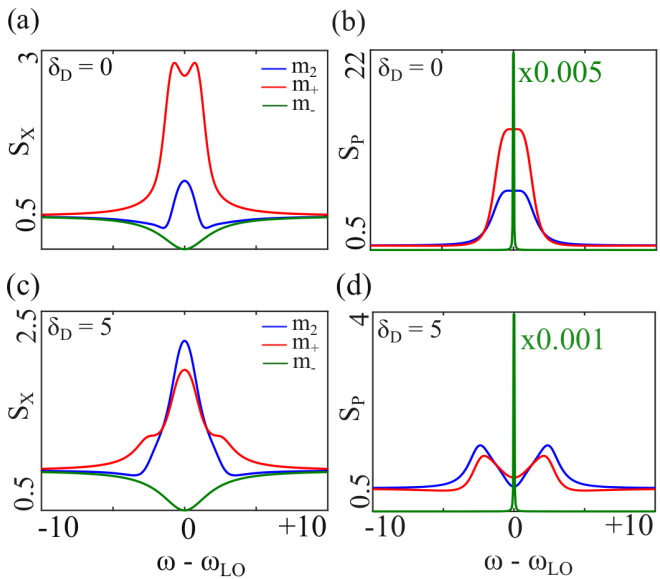}
\caption{~\label{fig:3-mode spectra} Output $X,P$ spectra of the modes in the (a),(b) harmonic cavity at $\Delta_0 = -3 , \Omega_0 = 1.85$, and (c),(d) anharmonic cavity at $\Delta_0 = 0 , \Omega_0 = 2$, calculated from the MF-Bogoliubov. In each panel the solid blue, red, and green lines correspond to the spectrum of the pumped ($\ket{m_2}$), symmetric ($\ket{m_+}$) and antisymmetric ($\ket{m_-}$) modes, respectively. Due to its divergence, the momentum of the antisymmetric mode is scaled down in panels (b),(d).}
\end{figure}

$\lambda_G = 0$ implies that in the local oscillators frame, $\ket{G}$ is a mode at $\omega=0$ with zero linewidth, i.e., an undamped excitation. To investigate the implications of this mode on quantum correlations, we employ Eq.~(\ref{eq:linearized fluctuations EoM}) to calculate the $XP$-quadrature spectra of the cavity output. 
Figure~\ref{fig:3-mode spectra} shows the quadrature correlations of the output $2^{nd}$-mode and $\ket{m_\pm} = m_1 \pm m_3$, i.e., the symmetric and antisymmetric superpositions of the two unpumped modes. Panels (a),(b) show the spectra of the harmonic cavity at $\Omega_0 = 1.85$, and panels (c),(d) show the same quantities for an anharmonic cavity at $\Omega_0 = 2$, which correspond to the point B within the MF LC-phase, and on the rising slope of the SDE/DM results in Fig.~\ref{fig:3-mode population-harmonic}(c) and Fig.~\ref{fig:3-mode population-anharmonic}(c).
Although the spectral features of the pumped and the symmetric mode depend on detail cavity features, the antisymmetric mode quadratures in harmonic and anharmonic cavities look alike (solid green lines in Fig.~\ref{fig:3-mode spectra}(c),(d)). While $S_{X_-}$ is unconditionally fully squeezed at the origin, the spectrum of its conjugated variable $S_{P_-}$ diverges. From Eq.~(\ref{eq:Goldstone mode}) it is clear that $S_{P_-}$ is indeed the spectrum of the gapless Goldstone mode. Since in the MF picture, this mode encounters no restoring force its fluctuation diverges. (The analytic form of the spectra and further can be found in Appendix~\ref{app:Covariance MF-Bog}.)

\begin{figure}[htbp]
\centering
\includegraphics[width=\linewidth]{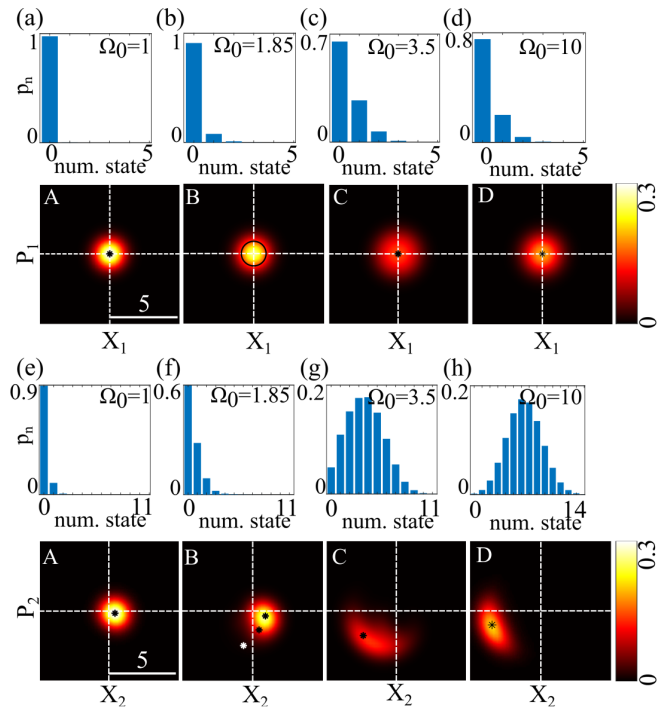}
\caption{~\label{fig:Wigner 3-mode Harmonic} Histograms of number state occupation probability $p_n$ and colormaps of the Wigner function of the (a)-(d) $1^{st}, 3^{rd}$-modes and (e)-(h) $2^{nd}$-mode, in a three-mode harmonic cavity when $\Delta_0 = -3 , V_0 = 1$ for different pumping rates $\Omega_0$ highlighted as (A,B,C,D) in Fig.~\ref{fig:3-mode population-harmonic}(c),(d). In each phase-space map the white dashed lines show the axes ($X=0,P=0$) in the $XP$-plane and black stars or circles correspond to the predicted MF.}
\end{figure}

\begin{figure}[htbp]
\centering
\includegraphics[width=\linewidth]{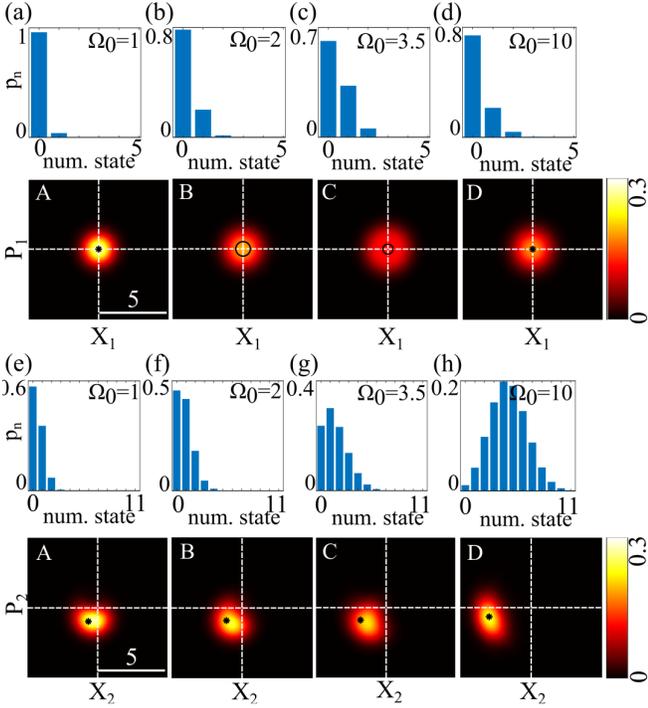}
\caption{~\label{fig:Wigner 3-mode anHarmonic} Histograms of number state occupation probability $p_n$ and colormaps of the Wigner function of the (a)-(d) $1^{st}, 3^{rd}$-modes and (e)-(h) $2^{nd}$-mode, in a three-mode anharmonic cavity when $\Delta_0 = 0, \delta_D = 5$ and at $V_0 = 1$ for different pumping rates $\Omega_0$ highlighted as (A,B,C,D) in Fig.~\ref{fig:3-mode population-anharmonic}(c),(d). In each phase-space map the white dashed lines show the axes ($X=0,P=0$) in the $XP$-plane and black stars or circles correspond to the predicted MF.}
\end{figure}

To examine the robustness of the Goldstone mode and the consequent unconditional squeezing, we employ the SDE to study the beyond-MF behavior of the cavity state. Figure~\ref{fig:Wigner 3-mode Harmonic} shows the number state occupation probability ($p_n$) and the Wigner function distribution of the harmonic cavity at four different pumping rates $\Omega_0 = 1,~ 1.85,~ 3.5,~ 10$ corresponding to (A,B,C,D) points in Fig.~\ref{fig:3-mode population-harmonic} at $V_0 = 1$, respectively. Panels (a)-(d) show these quantities for the $1^{st}$-mode and panels (e)-(h) show the ones for the $2^{nd}$-mode. As can be seen in all panels (a)-(d), distributions of the $1^{st},~3^{rd}$-modes are azimuthally symmetric independent of the pumping rate, which is consistent with the local U(1) symmetry of these two modes and their phase freedom, i.e., $\braket{\hat{a}_{1,3}} = 0$. 


Within the (A)-phase at low pumping rate and before the parametric threshold, MF predicts zero amplitude for the $1^{st},3^{rd}$ modes, while the $2^{nd}$ mode looks like a coherent state (Fig.~\ref{fig:Wigner 3-mode Harmonic}(a),(e)). 
As the pumping rate increases (point B in Fig.~\ref{fig:3-mode population-harmonic} (c),(d)), the system enters the LC-phase in which mode 2 has three stable fixed points, as shown with three stars in Fig.~\ref{fig:Wigner 3-mode Harmonic}(f), and the two unpumped modes acquire a finite population. The black circle in Fig.~\ref{fig:Wigner 3-mode Harmonic}(b) shows the loci of MF fixed points. 
For larger values of the pump, close to the upper threshold of the multi-stability region (point C in Fig.~\ref{fig:3-mode population-harmonic}(c),(d)), the systems transitions to the uniform (A)-phase again where the $2^{nd}$-mode attains a unique fixed point and the $1^{st},3^{rd}$-modes have zero MF. However, as can be seen in Fig.~\ref{fig:Wigner 3-mode Harmonic}(g) the cavity state is far from coherent due to the larger interaction at this photon number.

At even larger pumping rate shown in Fig.~\ref{fig:Wigner 3-mode Harmonic}(d),(h), corresponding to the point D in Fig.~\ref{fig:3-mode population-harmonic}(c),(d) (far within the (A)-phase), the $2^{nd}$ mode is a non-classical state whose phase-space distribution is reminiscent of the single-mode cavity at this regime (Fig.~\ref{fig:pump-only} of Appendix~\ref{app:single-mode}). Also it is worth mentioning that in spite of the similar symmetric distribution of the $1^{st},3^{rd}$ modes and their vanishing means, their variances clearly change as the system traverses through different phases.

For completeness, in Fig.~\ref{fig:Wigner 3-mode anHarmonic} we detail the state of the anharmonic cavity through its different phases at four pumping rates of $\Omega_0 =1,~ 2, ~ 3.5, ~ 10$ corresponding to (A,B,C,D) points in Fig.~\ref{fig:3-mode population-anharmonic}(c),(d). As can be seen the overall behavior of the cavity modes looks like that of the harmonic case, with the main distinction of always having one unique MF fixed point. 

To study the robustness of the Goldstone mode in the presence of quantum fluctuations, from SDE analysis we calculate the correlation and spectrum of $\hat{P}_-$ as 
\begin{align}\label{eq:minus-mode g1}
g^{(1)}(\tau) &= \braket{\lim_{t\to\infty}\hat{P}_-(t+\tau) \hat{P}_-(t)}, \\
S_{P_-}(\omega) &= \mathcal{F}\left(g^{(1)}(\tau) \right),
\end{align}
where $\hat{P}_- = i(\hat{a}_- - \hat{a}^\dagger_-)/\sqrt{2}$ is the momentum of $\ket{m_-}$-mode.

\begin{figure}[htbp]
\centering
\includegraphics[width=\linewidth]{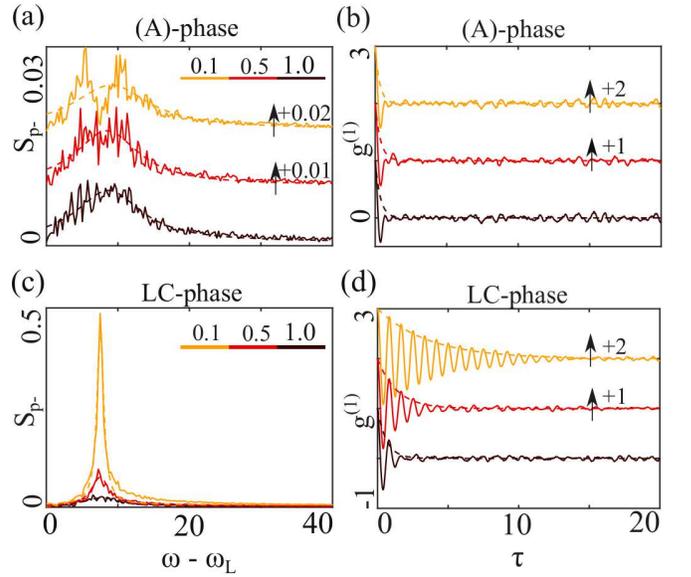}
\caption{~\label{fig:goldstone mode spec} SDE calculations of the (a),(c) spectral density in the laser frame and (b),(d) delayed temporal correlation of $P$-quadrature of $\ket{m_-}$ mode in an anharmonic cavity. The upper row shows the behavior in the (A)-phase and the lower row shows the ones with the MF LC-phase. The interaction is changed from $V_0 = 0.1$ to $V_0 = 1.0$, yellow to red to brown, respectively. The dashed lines show the Lorentzian fit in (a),(c) and the exponential fits in (c),(d).}
\end{figure}

The results are shown in Fig.~\ref{fig:goldstone mode spec}  when the interaction $V_0$ is increased from 0.1 to 1 (brown to yellow lines). Panels (a),(b) are the spectra and correlations in the (A)-phase while (c),(d) are within the (C)-phase where LC is predicted by MF. For direct comparison with LC oscillations of Fig.~\ref{fig:LC}(b) and highlighting $\omega_{LC}$, the spectral densities in (a),(c) are shown in the laser ($\omega_L$) rather than the local frame ($\omega_{LO}$). 
Defining a dimension-less parameter $N$ where $V_0/N \rightarrow 0^+$ is the TD limit, the pumping rate is scaled by $\sqrt{N}$, so that $V \Omega^2$ is kept fixed. 

As can be seen in Fig.~\ref{fig:goldstone mode spec}(a),(b), the observables are almost unchanged when the system is in the (A)-phase, where MF predicts zero-photon number in $\ket{m_-}$. From Fig.~\ref{fig:goldstone mode spec}(a) we can see that the linewidth of this mode is large and the spectral density is very small (note that the lines for $V_0 = 0.5, ~ 0.1$ are shifted upwards to clarify things better). Similarly, the temporal behavior in panel (b) shows a short correlation time. 

On the contrary, when the system transitions to the MF LC-phase by virtue of increasing the pumping rate, the spectral densities shown in Fig.~\ref{fig:goldstone mode spec}(c) increase and an apparent resonance feature appears that becomes more prominent at weaker interaction closer to the TD limit hence, the validity range of MF.
Similarly the temporal correlations in panel (d) show prolonged coherence times that increases at weaker interaction. To quantify these features better we fit a Lorentzian lineshape with the following form to $S_{P_-}(\omega)$ 
\begin{equation}~\label{eq:lor. fit}
    L(\omega) = \frac{a}{(\omega - \omega_{peak})^2 + \Gamma^2} + c
\end{equation}
The fits are shown with dashed lines in Fig.~\ref{fig:goldstone mode spec}(a),(c) and the center and linewidth fit parameters are presented in table~\ref{tab: Goldstone mode Lorentzian}.
Within the (A)-phase, $\omega_{peak},\Gamma$ slightly changes with changing the interaction $V_0$. Throughout the LC-phase on the other hands, $\omega_{peak} \approx 7.5$, i.e., the LC oscillation frequency $\omega_{LC}$ in Fig.~\ref{fig:LC}(b). Moreover, starting from a narrow resonance ($\Gamma \approx 0.4$) at weak interaction (large $N$), the linewidth clearly increases ($\Gamma \approx 3.2$) by increasing the interaction (small $N$). Similar values were obtained by fitting the correlation functions with exponential functions, i.e. dashed lines in Fig.~\ref{fig:goldstone mode spec}(b),(d), independently. 

\begin{table}~\label{tab: Goldstone mode Lorentzian}
\begin{tabular}{ |c|c|c|c| } 
\hline
 &                        $V_0 = 0.1$ & $V_0 = 0.5$  & $V_0 = 1.0$ \\
\hline
$\omega_{peak}$ (A)    & 8.7   & 8.5  & 9  \\
\hline
$\omega_{peak}$ (LC)       & 7.5   & 7.7  & 7.9 \\
\hline
$\Gamma$ (A)           & 5.6   & 5.4  & 6.5 \\
\hline
$\Gamma$ (LC)              & 0.4  & 1.7  & 3.2 \\
\hline
\end{tabular}
\caption{The Lorentzian fit parameters to the spectral density of $P_-$quadrature within the MF (A)- and LC-phase as in Fig.~\ref{fig:goldstone mode spec}(a),(c).}
\end{table}
\begin{figure}[htbp]
\centering
\includegraphics[width=\linewidth]{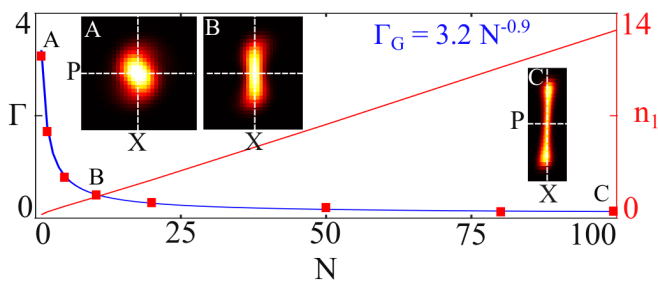}
\caption{~\label{fig:goldstone mode decay} The linewidth of $P_-$quadrature, i.e. the Goldstone mode, within the MF LC-phase as a function of dimensionless parameter $N$. The red squares are the SDE calculation results while the solid blue line is a power-law fit to the data. The solid red line shows the number of particles in this mode (right axis). The inset colormaps show the TW distribution of $\ket{m_-}$-mode at a couple of interaction strengths.}
\end{figure}
As a final remark we study the behavior of $P_-$quadrature linewidth within the whole quantum to TD limit, corresponding to the small and large $N$, respectively. The results depicted in Fig.~\ref{fig:goldstone mode decay} with red squares. The solid red line is the number of particles in this mode (right y-axis), and the the solid blue line is a power-law fit to the data, indicating the linewidth narrowing scales as $N^{-0.9}$. In other words, while the gapless Goldstone mode picture at TD limit (kernel of MF-Bogoliubov matrix) corroborates well with a small $\Gamma \approx 0$ of $P_-$quadrature, approaching the quantum limit the decay rate notably increases due to the phase diffusion.
It is worth comparing this tendency with $N^{-1}$ behavior, i.e. the Shallow-Towens laser linewidth scaling~\cite{Haken1984}. 

To investigate the $\ket{m_-}$-mode noise spectra as well, we show the Wigner function distribution of this mode at a few different interaction points. As can be seen at larger $N$ (point C), hence the weaker interaction, the phase-space distribution resembles the one of a number-squeezed state. However, upon increasing the interaction (points A,B) the squeezing decreases. 
This clearly confirms the phase diffusion effect in reducing the coherence time of the generated pairs. Besides, this effect becomes more dominant deep into the quantum range where the fluctuations should not be ignored. 
\section{Conclusion}
Exploring dissipative phase transitions is one of the important topics of open quantum systems. There, the interplay between dissipation, drive, and interaction can lead to a rich testbed to investigate dynamics of many-body systems far from their equilibrium.  In this article, we theoretically investigate the first- and second-order quantum dissipative phase transitions of in a three-mode cavity with intra- and inter-modal two-body interaction as a prototypical model. We showed the emergence of a MF limit-cycle phase where the local U(1) symmetry and the TTS of the Liouvillian are spontaneously broken. We explained the connection between this phase and the Goldstone mode well-studied in the TD limit. By employing the Wigner function formalism hence, properly including the quantum noise, we showed the breakdown of MF predictions within the quantum regime. Within this range, fluctuations notably limit the coherence time of the Goldstone mode due to the phase diffusion. 

Concerning the experimental realizations, the model and the results are applicable to a wide variety of driven-dissipative interacting bosonic platforms, including circuit-QED, semiconductor excitons, and multi-mode cavities with cold-atoms~\cite{Jia2018, Vaidya2018}, where the figure of merit $V_0/\gamma$ can be tuned, properly. It is also interesting to explore the feasibility of using such platforms in creating non-Gaussian states as an instrumental ingredient for quantum information protocols based on continuous variable entanglement and photonic quantum logic gates~\cite{Braunstein2005,Santori2014,Liu2017,Zhang2017}.

\section*{acknowledgement}
The authors thank Wolfgang Schleich, Hans-Peter B\"uchler, Jan Kumlin, and Jens Hertkorn for insightful discussions. The invaluable IT support from Daniel Weller is greatly acknowledged. H. A. acknowledges the financial supports from IQST Young Researchers grant and the Eliteprogram award of Baden-W\"urttemberg Stiftung. I. C. acknowledges financial support from the European Union FET-Open grant ``MIR-BOSE'' (n. 737017), from the H2020-FETFLAG-2018-2020 project "PhoQuS" (n.820392), and from the Provincia Autonoma di Trento. 

\bibliography{ref}

\newpage
\appendix

\section{Covariance Matrix from MF Bogoliubov}~\label{app:Covariance MF-Bog}
As described in the main text, the equations of motion for the fluctuation operators $[\hat{b}(t)]$ are given by the linearized Eq.~(\ref{eq:linearized fluctuations EoM}). From this equation we can determine directly the Fourier transform of fluctuation operator $\hat{B}$ for the inside-cavity fields as 
\begin{equation}~\label{eq:fluctuation spectra}
    \hat{B}(\omega) = -\left(i\omega I + M\right)^{-1} D^{1/2}\hat{\Xi}(\omega),
\end{equation}
where $\omega$ is the frequency in the local oscillator ($\omega_{LO}$) rotated frame. For dynamically stable solutions, i.e., $M$ being negative-definite, the above solution always exists. We define the following covariance matrix spectrum with entries as in Eq.~(\ref{eq:spectral response})
\begin{widetext}
\begin{equation*}~\label{eq: covariance matrix}
    C_B^{(in)}(\omega) = \braket{\hat{B}(\omega) \hat{B}^\dagger (-\omega)} = \braket{\hat{B}(\omega) \hat{B}(\omega)^\dagger}  
    = \left(i\omega I + M\right)^{-1} D^{1/2} \braket{\hat{\Xi}(\omega) \hat{\Xi}^\dagger(-\omega)} D^{1/2} \left(i\omega I + M\right)^{-1^\dagger},
\end{equation*}
\end{widetext}
where the superscript $(in)$ refers to the inside-cavity fields, and the subscript $B$ emphasizes the operators. $\braket{\hat{\Xi}(\omega) \hat{\Xi}^\dagger(-\omega)}$ is the noise spectral density, solely dependent on bath features, e.g. thermal photons density in our case. The coupling rate of the inside-cavity dynamics with the surrounding bath is captured via matrix $D$. Other detailed information about the bare cavity modes, pumping, and interactions are in matrix $M$. At any stable MF, $M$ is a negative matrix so $C_B^{(in)}(\omega)$ is a well-defined quantity over the whole spectrum except $\omega = 0$, in case $M$ has a kernel.

Employing the input-output formalism the output-field covariance matrix can be determined directly from $C^{(in)}(\omega)$. For a single-sided cavity we have
\begin{equation}~\label{eq: input-output formalism}
    \hat{B}_{out} = D^{1/2} \hat{B}_{in} - \hat{\Xi}.
\end{equation}
Which leads to the following covariance matrix for the output field as
\begin{widetext}~\label{eq: output covariance matrix}
\begin{multline*}
    C_B^{(out)}(\omega) = \braket{\hat{B}^{(out)}(\omega) \hat{B}^{(out)}(\omega)^\dagger}  
    = \left(I + D^{1/2}\left(i\omega I + M\right)^{-1} D^{1/2}\right) \braket{\hat{\Xi}(\omega) \hat{\Xi}^\dagger(-\omega)} \left(I + D^{1/2} \left(i\omega I + M\right)^{-1}D^{1/2} \right)^\dagger.
\end{multline*}
\end{widetext}
From Eq.~(\ref{eq:mean-field equations}) it is straightforward to shows that MFs satisfy the following equations for $\alpha_m \rightarrow \alpha_m e^{i\phi_m}$ and $\Phi_0  = 2\phi_2 - \phi_1 - \phi_3 $, $\Phi_p = \phi_p - \phi_2$
\begin{widetext}~\label{eq:polar MF}
\begin{align*}
    \left(\frac{\alpha_1}{\alpha_3}\right)^2 &= \frac{\gamma_3}{\gamma_1} ~ , ~
    \sin(\Phi_0) = \frac{\sqrt{\gamma_1 \gamma_3}}{V_0 \alpha_2^2} ~ , ~
    \Omega_0 \sin(\Phi_p) = \alpha_2 \left(1 + 2\alpha_2^2 \alpha_1 \alpha_3 \sin(\Phi_0) \right)
    \\
    \Tilde{\Delta}_1 &= \frac{\gamma_1}{\gamma_1 + \gamma_3}(2\Delta_2 - \delta_D) + \frac{V_0 \alpha_2^2}{\gamma_2(\gamma_1 + \gamma_3)}\left(2(\gamma_1 - \gamma_3) + 2(\gamma_1 \alpha_1^2 - \gamma_3 \alpha_3^2) + (\gamma_1 \alpha_3^2 - \gamma_3 \alpha_1^2) \right) \\
    \Tilde{\Delta}_3 &= \frac{\gamma_3}{\gamma_1 + \gamma_3}(2\Delta_2 - \delta_D) + \frac{V_0 \alpha_2^2}{\gamma_2(\gamma_1 + \gamma_3)}\left(2(\gamma_3 - \gamma_1) + 2(\gamma_3 \alpha_3^2 - \gamma_1 \alpha_1^2) + (\gamma_3 \alpha_1^2 - \gamma_1 \alpha_3^2) \right).
\end{align*}
\end{widetext}
Note that $\Tilde{\Delta}_{1,3}$ are the renormalized detunings after extracting the LC oscillations. They therefore depend on other system parameters. From these equations it is clear that if $\gamma_1 = \gamma_3$ then the field amplitudes and their renormalized detuning are the same. Moreover, $\Delta_{1,3}$ becomes MF-independent solely dependent on mode frequencies and the pumping rate. 
The difference between $\Delta_{1,3}$ in Eq.~(\ref{eq:Heisenberg-Langevin}), i.e. the detuning in the laser rotated frame, and $\Tilde{\Delta}_{1,3}$ in the above equation is the LC oscillations depicted in Fig.~\ref{fig:LC} of the main text. It is straightforward to check that the LC oscillations have the following frequency
\begin{equation}~\label{eq: LC freq.}
    \Tilde{\Delta}_{1,3} - \Delta_{1,3} = \pm \omega_{LC} = \pm \frac{\omega_3 - \omega_1}{2}.
\end{equation}
Note that the last equation is indeed the same as local U(1)-symmetry of Eq.(\ref{eq:local U(1)-sym}).

To study the squeezing it is often more suitable to investigate the behavior of field quadratures, related to field operators $\hat{B}$ via a unitary transformation $U$ as $\hat{XP} = U \hat{B}$. Their covariance matrix reads as 
\begin{equation}~\label{eq: XP covariance}
    C_{XP}(\omega) = \braket{\hat{XP}(\omega) \hat{XP}(\omega)^\dagger} = U C_B(\omega) U^\dagger.
\end{equation}
For the three-mode cavity investigated in this work, we define new modes as the rotation ($m_2$) and symmetric and asymmetric superposition of the cavity modes ($m_\pm$) as
\begin{equation}~\label{eq:new cavity modes}
    \hat{A}_2 = \hat{a}_2 e^{i\phi_2} ~, ~ \hat{A}_\pm = \frac{\alpha_1 \hat{a}_1 e^{i\phi_1} \pm \alpha_3 \hat{a}_3 e^{i\phi_3}}{\sqrt{\alpha_1^2 + \alpha_3^2}}.
\end{equation}
The generalized quadratures of these modes are defined as follow
\begin{equation}~\label{eq:generalized XP}
\begin{aligned}
    \hat{X}_2(\theta_2) &= \frac{\hat{A}_2e^{i\theta_2} + \hat{A}_2^\dagger e^{-i\theta_2 }}{\sqrt{2}} \\
    \hat{X}_\pm(\theta_{1,3}) &= \frac{\hat{X}_1(\theta_1) \pm  \hat{X}_3(\theta_3)}{\sqrt{2}} .
\end{aligned}
\end{equation}
The momentum quadratures are defined, similarly. Note that $\hat{P}_-$ is the operator associated with the Goldstone mode $\ket{G}$ in Eq.(\ref{eq:Goldstone mode}).

A direct calculation of $M$ for $XP$-operators indicate that for $(\theta_1 + \phi_1) = (\theta_3 + \phi_3)$, $M_{XP} = M_{4 \times 4} \bigoplus M_{2 \times 2}$, decoupling the dynamics of the $m_2 , m_+$-modes from the $m_-$. For $\theta_2 = -\phi_2$ the Bogoliubov matrix has the following form 
\begin{widetext}~\label{eq: 3-mode Bog. M}
\begin{multline*}
M_{XP} = \\
\begin{bmatrix}
\frac{\Omega_0}{\alpha_2}\sin{\phi_2} & 4V_0\alpha_1^2 \cos{\Phi_0} + \frac{\Omega_0}{\alpha_2}\cos{\phi_2} & -2\sqrt{2}V_0 \alpha_1 \alpha_2 \sin{\Phi_0} & -2\sqrt{2}V_0 \alpha_1 \alpha_2 \cos{\Phi_0} & 0 & 0 \\
2V_0 \alpha_2^2 - \frac{\Omega_0}{\alpha_2}\cos{\phi_2} & 4V_0\alpha_1^2 \sin{\Phi_0} + \frac{\Omega_0}{\alpha_2}\sin{\phi_2} & 2\sqrt{2}V_0 \alpha_1 \alpha_2 \left(2 + \cos{\Phi_0} \right) & -2\sqrt{2}V_0 \alpha_1 \alpha_2 \sin{\Phi_0} & 0 & 0 \\
2\sqrt{2}V_0 \alpha_1 \alpha_2 \sin{\Phi_0} & -2\sqrt{2}V_0 \alpha_1 \alpha_2 \cos{\Phi_0} & 0 & 2V_0 \alpha_2^2 \cos{\Phi_0} & 0 & 0 \\
2\sqrt{2}V_0 \alpha_1 \alpha_2 \left(2 + \cos{\Phi_0}\right) & 2\sqrt{2}V_0 \alpha_1 \alpha_2 \sin{\Phi_0} & 6 V_0 \alpha_1^2 & -2 V_0 \alpha_2^2 \sin{\Phi_0} & 0 & 0 \\
0 & 0 & 0 & 0 & -2V_0 \alpha_2^2 \sin{\Phi_0} & 0\\
0 & 0 & 0 & 0 & -2V_0\left(\alpha_1^2 + \alpha_2^2 \cos{\Phi_0} \right) & 0
\end{bmatrix}
\end{multline*}
\end{widetext}
Clear $M$ has a kernel along $P_-$, hence a gap-less mode without any further dynamics. 

The output spectrum of this mode can be directly obtained from $C_B^{(out)}(\omega)$ in Eq.~(\ref{eq: 3-mode Bog. M}). For brevity we define matrix $N$ as 
\begin{equation}~\label{eq:N-matrix}
    N = I + D^{1/2}(i\omega I + M)^{-1} D^{1/2}.
\end{equation}
From Eq.~(\ref{eq: output covariance matrix}) we get
\begin{widetext}~\label{eq: XP covariance}
\begin{align*}
C_{XP}^{(out)}(\omega) &= U \left[N (U^\dagger U) \braket{\Xi(\omega) \Xi^\dagger(-\omega)} (U^\dagger U) N^\dagger\right] U^\dagger = \left(U N U^\dagger\right) \left(U \braket{\Xi(\omega) \Xi^\dagger(-\omega)} U^\dagger \right) \left(U N U^\dagger\right)^\dagger 
    \\
U N U^\dagger &= U \left(I + D^{1/2} (i\omega I + M)^{-1} D^{1/2} \right) U^\dagger = I + D^{1/2} U (i\omega I + M)^{-1} U^\dagger D^{1/2} \\ \nonumber
&= I + D^{1/2} \left[U (i\omega I + M) U ^{-1}\right]^{-1}  D^{1/2} =
I + D^{1/2} (i\omega I + M_{XP})^{-1} D^{1/2},
\end{align*}
\end{widetext}
where $M_{XP} = U M U^{-1}$ is the Bogoliubov matrix of the generalized rotated quadratures. 
Note that in the above equation we implicitly assumed identical losses for all modes i.e., $D^{1/2} = \sqrt{2\gamma_0} I$. 
Finally, the quadrature spectra of $m_-$-mode will be obtained as

\begin{widetext}~\label{eq:Sx,Sp}
\begin{align}
    S_{X_-}(\omega) &= \frac{1}{2}\left(\frac{(M_{55} + 2\gamma_0)^2 + \omega^2}{M_{55}^2 + \omega^2} \right) =  \frac{1}{2}\left(\frac{\omega^2}{M_{55}^2 + \omega^2} \right) \\
S_{P_-}(\omega) &= \frac{1}{2}\left(1 + \frac{4\gamma_0 [\gamma_0(\omega^2 + M_{55}^2 + M_{65}^2) +  M_{65}(2\gamma_0 + M_{55})]}{\omega^2 (\omega^2 + M_{55}^2)} \right) = \frac{1}{2}\left(1 + \frac{4\gamma_0^2(\omega^2 + M_{55}^2 + M_{65}^2) }{\omega^2 (\omega^2 + M_{55}^2)} \right),
\end{align}
\end{widetext}
where we used the expression for $\Phi_0$ from Eq.~(\ref{eq:polar MF}) to simplify the final form of the spectrum.
These two spectra have simple interpretations; first, they show that at local oscillator frame $\braket{\Delta^2 P_-}$ diverges while the $\braket{\Delta^2 X_-}$ vanishes. Moreover, for all frequencies the momentum quadrature is always above SQL ($\ge 0.5$), and $S_{X_-} \le 0.5$ below SQL, indicating the squeezing of this quadrature. Both quantities asymptotically approach SQL at large frequencies, as expected for the asymptotic vacuum noise.

\section{Wigner representation}~\label{app:Wigner func.}
The Wigner representation of the density matrix $\hat{\rho}$ in the complex plane can be derived from Eq.~(\ref{eq:master equation}) by assigning c-numbers $\alpha^\pm _m$ for each degrees of freedom and using the following relations to replace the operator algebras with calculus ones on analytic function $W$. A detailed explanation of the Weyl transformation and Wigner function representation can be found in~\cite{Carmichael1991,Steel1998,Gardiner2004,Wiseman2011}, 
\begin{align}~\label{eq:Wigner transform}
    \hat{a}_m \hat{\rho} \rightarrow \left(\alpha_m^- + \frac{1}{2} \frac{\partial}{\partial \alpha_m^+} \right) W(\alpha_m^\pm;t), \\ \nonumber
    \hat{a}^\dagger_m \hat{\rho} \rightarrow \left(\alpha_m^+ - \frac{1}{2} \frac{\partial}{\partial \alpha_m^-} \right) W(\alpha_m^\pm;t), \\ \nonumber
\end{align}

\begin{align}~\label{Wigner transform2}
    \hat{\rho} \hat{a}_m \rightarrow \left(\alpha_m^- - \frac{1}{2} \frac{\partial}{\partial \alpha_m^+} \right) W(\alpha_m^\pm;t), \\ \nonumber
     \hat{\rho} \hat{a}^\dagger_m \rightarrow \left(\alpha_m^+ + \frac{1}{2} \frac{\partial}{\partial \alpha_m^-} \right) W(\alpha_m^\pm;t).
\end{align}
Different terms of the master equation can be replaced with their equivalent form in terms of $W(\alpha_m^\pm)$, where $\alpha_m^\pm = (\alpha_m^\mp)^*$. For the bare-cavity dynamics as $\omega_m \hat{a}_m^\dagger \hat{a}_m$ we have
\begin{align}~\label{eq:free mode}
     \frac{\partial}{\partial \alpha_m^-} \left((i\omega_m + \gamma_m) \alpha_m^-\right) + \frac{\gamma_m}{2}\frac{\partial^2}{\partial \alpha_m^- \partial\alpha_m^+} + c.c.^+ ,
\end{align}
where $c.c^+$ in these equations represents the complex conjugate.

The self-phase modulation (SPM) as $ \hat{a}_m^{\dagger^2} \hat{a}_m^2$ gets transformed to
\begin{align}\label{eq:SPM}
     \frac{\partial}{\partial \alpha_m^-} \left(\alpha_m^{-^2}\alpha_m^+ - \alpha_m^{-}\right) +   \frac{1}{4}\frac{\partial^3}{\partial \alpha_m^- \partial^2\alpha_m^{+}} \alpha_m^+ + c.c.^+
\end{align}
The cross-phase modulation (XPM) $\hat{a}_m^\dagger \hat{a}_n^\dagger \hat{a}_n\hat{a}_m, n\not=m$ will be given as
\begin{align}~\label{eq:XPM}
     \frac{\partial}{\partial \alpha_m^-} \left(\alpha_m^- \alpha_n^- \alpha_n^+ - \frac{\alpha_m^-}{2} \right) +
     \frac{\partial}{\partial \alpha_n^-} \left(\alpha_n^- \alpha_m^- \alpha_m^+ - \frac{\alpha_n^-}{2} \right)  \\ \nonumber
     -\frac{\partial^3}{\partial \alpha_m^- \partial\alpha_m^+ \partial\alpha_n^-} \left(\frac{\alpha_n^-}{2} \right) - 
     \frac{\partial^3}{\partial \alpha_n^- \partial\alpha_n^+ \partial\alpha_m^-} \left(\frac{\alpha_m^-}{2} \right) + 
     c.c.^+
\end{align}
And finally the exchange term as $\hat{a}_1^\dagger \hat{a}_3^\dagger \hat{a}_2^2 + H.C.$ gets the following form 
\begin{widetext}~\label{eq: Exch}
\begin{align*}
    \frac{\partial}{\partial \alpha_1^-} \left(\alpha_2^{-^2}\alpha_3^+ \right) + 
    \frac{\partial}{\partial \alpha_3^-} \left(\alpha_2^{-^2}\alpha_1^+ \right) +
    \frac{\partial}{\partial \alpha_2^-} \left(2 \alpha_1^{-}\alpha_3^- \alpha_2^+ \right) + c.c.^+ \\
     -\frac{\partial^3}{\partial \alpha_3^+ \partial^2\alpha_2^{-}}\left(\frac{\alpha_1^-}{4}\right)
     -\frac{\partial^3}{\partial \alpha_1^- \partial\alpha_3^- \partial\alpha_2^+}\left(\frac{\alpha_2^-}{2}\right)
    -\frac{\partial^3}{\partial \alpha_1^+ \partial^2\alpha_2^{-}}\left(\frac{\alpha_3^-}{4}\right)
\end{align*}
\end{widetext}
When inserted into Eq.~(\ref{eq:master equation}), on can determine the equation of motion for $W(\vec{\alpha};t)$ as
\begin{equation}~\label{eq:Fokker-Planck}
    \frac{\partial}{\partial t} W(\alpha_m^\pm;t) = \mathcal{L}_W[W(\alpha_m^\pm;t)].
\end{equation}
In the above equation $\mathcal{L}_W$ is a 3$^{rd}$-order differential operator acting on the analytic function $W(\vec{\alpha};t)$ and is equivalent to the super-operator $\mathcal{L}$ acting on the density matrix $\hat{\rho}$. 
If the 3$^{rd}$-order derivatives in $\mathcal{L}_W[.]$ are ignored, the resulting truncated Wigner function $W$ turns into a Fokker-Planck equation with the following general form~\cite{Berg2009}
\begin{equation}~\label{FP-truncated Wigner}
    \frac{\partial}{\partial t}W(\vec{\alpha};t) \approx -\sum_m \partial_{\alpha_m}(A_m W) + \frac{1}{2} \sum_{m,m'} \partial^2 _{\alpha_m \alpha_m'} (D D^T W),
\end{equation}
where $A_m , D$ represent the drift and diffusion matrices in a stochastic process, respectively.

\section{Summary of DPT in a Single-Mode Cavity}~\label{app:single-mode}
\begin{figure}[htbp]
\centering
\includegraphics[width=\linewidth]{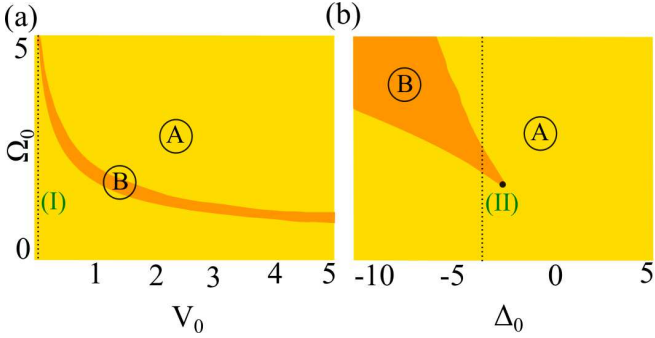}
\caption{\label{fig:pump-only PT} MF Dissipative phase diagram of a single-mode cavity as a function of (a) the interaction strength $V_0$ and (b) the laser detuning $\Delta_0$. In each panel the yellow (A) and orange (B) regions correspond to one and two (bi-stability) fixed points for the pumped mode, respectively. In (a) the detuning is fixed at $\Delta_0 = -3$ and in (b) the interaction strength has the constant value $V_0 = 1$. The dotted vertical lines [labelled (I) and (II)] at $V_0=0.1$ and $\Delta_0=-3$ indicate the the cuts through the phase diagram studied in subsequent figures.}
\end{figure}
\subsubsection{First-order DPT in a single-mode cavity}
Starting from Eq.~(\ref{eq:mean-field equations}) we can drive the following equation for the mean photon number in the cavity mode as
\begin{equation}~\label{eq: single-mode cubic MF}
    V_0^2 n^3 + 2\Delta_0 V_0 n^2 + (\Delta_0^2 + 1)n = \Omega_0^2
\end{equation}
where $n = |\alpha_2|^2$ is the photon number and all the rates are normalized to $\gamma_0$, as usual. 
This cubic equation can be solved, exactly to give three values for $n$ at each $\Omega_0$. However, $n$ being a real positive quantity, imposes additional constraints for having a physical results. The discriminant of this cubic equation reads as
\begin{equation}~\label{eq: discreminant}
    \Delta = -V_0^2 \left(27V_0^2 \Omega_0^4 + 4V_0\Delta_0 \Omega_0^2 (9 + \Delta_0^2) + 4(1+\Delta_0^2)^2 \right)
\end{equation}
For a repulsive interaction i.e. $V_0 \ge 0$ and for a red-detuned coherent excitation $\Delta_0 \ge 0$, the discriminant $\Delta \le 0$, hence the system always has a single real solution which is positive in this case $\Omega_0^2/V_0^2 \ge 0$. Following a dynamical stability analysis one can show that this solution is stable as well hence, it is the solution of the non-linear cavity, as well as depicted in the phase diagram of Fig.~\ref{fig:pump-only PT}(b). For a blue-detuned excitation $\Delta_0 \le 0$, the same argument holds as while as the term in parenthesis remains positive. For each detuning $\Delta_0$, this puts an upper and lower bound on the pumping rate $\Omega_0$. These are the boundaries between the yellow and orange regions in Fig.~\ref{fig:pump-only PT}.

\begin{figure}[htbp]
\centering
\includegraphics[width=\linewidth]{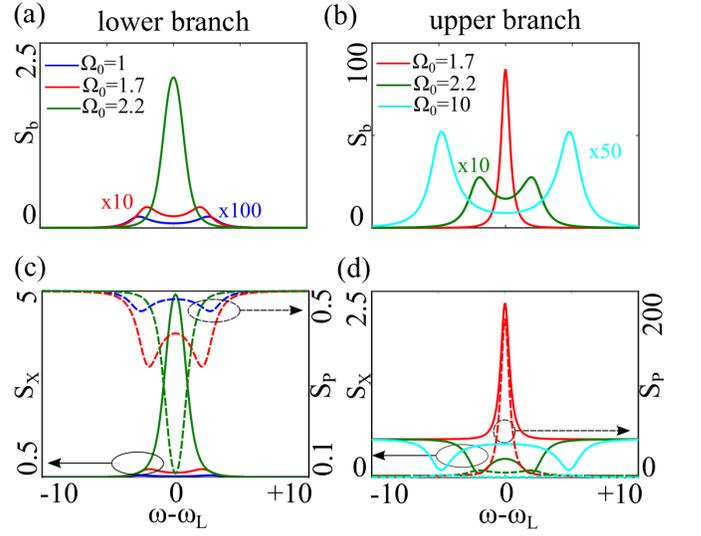}
\caption{\label{fig:pump-only-spectra} Inelastic output intensity spectrum in the laser reference frame, i.e., ($\omega - \omega_L$) when the pumping rate is increased on the MF (a) lower and (b) upper branches at $\Delta_0 = -3 , V_0 = 1$. (c) and (d) show their corresponding $X,~P$ variance spectra calculated from Eq.~(\ref{fluctuation spectra}). Solid lines show $X$ spectra and dashed lines are their corresponding $P$ spectra.}
\end{figure}

These two threshold pumping values lead to two different values for $n$ in Eq.~(\ref{eq: single-mode cubic MF}), hence an abrupt change in the particle number as shown in Fig.~\ref{fig:pump-only PT}. For pumping rates in between, Eq.~(\ref{eq: discreminant}) leads to $\Delta \ge 0$, hence three different real solutions for the cubic Eq.~(\ref{eq: single-mode cubic MF}) exist. Moreover, these roots are positive hence indeed they can be physical solutions for $n$. However, the dynamical stability analysis indicates that only particle numbers satisfying $(V_0 n + \Delta_0)(3V_0 n + \Delta_0) \ge 0$ are stable MF solutions. Since the upper and lower branches should remain continuous, the intermediate solution for $n$ within the multi-stability region is not acceptable, which means the orange multi-stable region in Fig.~\ref{fig:phase transition}(a),(b) is a bi-stable phase. Physically the instability of this solution is due to the divergence of parametrically-generated side peaks shown in Fig.~\ref{fig:pump-only-spectra}(a),(b).

Figure~\ref{fig:pump-only PT} (a),(b) shows the MF-DPT of a single-mode cavity as a function of the interaction strength ($V_0$), the detuning ($\Delta_0$), and the pumping rate ($\Omega_0$). In each panel the yellow region shows the single-solution conditions while the orange ones correspond to the parameter ranges where the system has two different stable solutions (bi-stability region).

\begin{figure}[htbp]
\centering
\includegraphics[width=\linewidth]{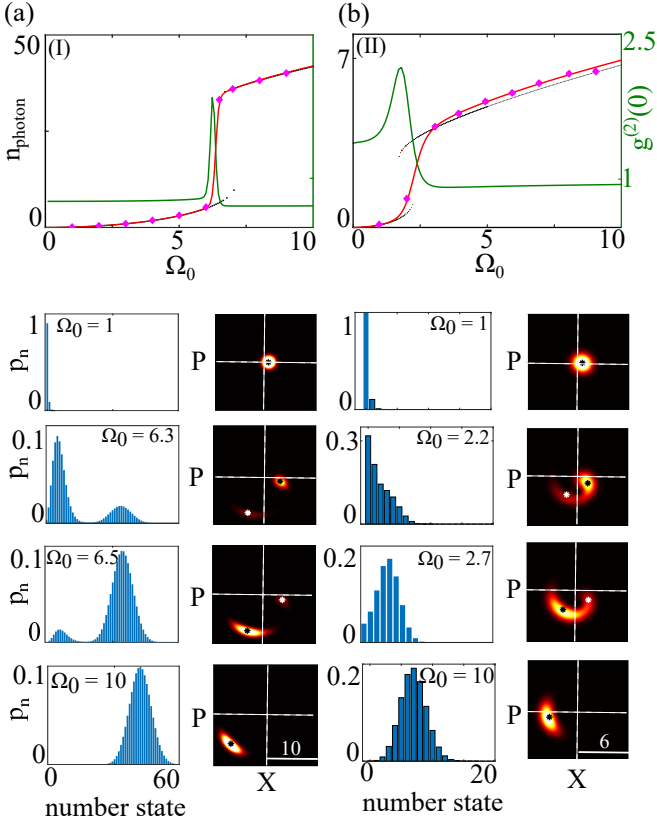}
\caption{\label{fig:pump-only} Number of photons as a function of pumping rate ($\Omega_0$) in a single-mode cavity with two-body interaction strength of (a) $V_0 = 0.1$ (dotted line (I) in Fig.~\ref{fig:pump-only PT}(a)) and (b) $V_0 = 1$ (dotted line (II) in Fig.~\ref{fig:pump-only PT}(b)). The black dots show the mean-field result. The solid red line is the full density matrix result, and the purple diamonds represent the SDE results. The solid green line shows the intensity fluctuation $g^{(2)}(0)$. For each interaction strength the blue histograms show the number state occupation probability $p_n$ at different pumping rates, before the bi-stability (first row), within the bi-stability (two middle rows) and after the bi-stability (last row). In each case, the colormaps show the corresponding Wigner function distribution in the $XP$-phase space. The white dashed lines show the axes ($X=0,~P=0$) in that plane. Also the black or white stars in those panels indicate the predicted MFs. We use $\Delta_0 = -3$ in all cases.}
\end{figure}
To better understand the system behavior in different phases we investigate the dependence of the cavity photon number on the pumping rate $\Omega_0$, at a fixed detuning $\Delta_0 = -3$. The results obtained from the three different methods are compared in Fig.~\ref{fig:pump-only} (a),(b) for weak ($V_0 = 0.1$ corresponding to the dotted line (I) in Fig.~\ref{fig:phase transition}(a)) and strong interaction ($V_0 = 1$ corresponding to the dotted line (II) in Fig.~\ref{fig:pump-only PT}(b)), respectively. 

As can be seen, at low pumping rate and for both weak and strong interaction $V_0$, a displaced thermal state emerges inside the cavity. The effective temperature $T$ increases with interaction strength starting from from 0 at $V_0 = 0$. Notice the larger deviation of $g^{(2)}(0)$ from unity when the interaction is increased as in Fig.~\ref{fig:pump-only}(a) to (b). Within this range there is a good agreement between all three approaches.

As the pumping rate increases, however, the MF predicts a bi-stable behavior corresponding to a \emph{first-order} phase transition, while both DM and SDE give a unique solution. In the MF picture, once the system reaches any of the stable fixed points the dynamics stops and the system stays there forever. Quantum mechanically, however, due to fluctuations these solutions are only \emph{meta-stable} states and the system can switch between them. 
The signature of the quantum tunneling/switching between these states can be clearly observed in an increase of the intensity fluctuation $g^{2}(0)$ within the bi-stability region as depicted in solid green line in Fig.~\ref{fig:pump-only}(a),(b). Its deviation from unity outside the bi-stability range is another apparent deviation from MF. 

To explore the tunneling phenomenon further, we compare the behavior of the system at weak (Fig.~\ref{fig:pump-only}(a)) and strong (Fig.~\ref{fig:pump-only}(b)) interaction as a function of the pumping rate.
As can be seen by decreasing the interaction, the onset of bi-stability, the average number of cavity photons, and the difference between stable branches all increase. At weaker interactions, hence larger particle number, the fluctuations can be neglected and the quantum mechanical predictions approach the stable MF solutions. In the tunneling picture it can be understood as an increase of the barrier height at weaker interactions hence, rare tunneling events. This rate will be noticeably decreased upon increasing the interaction strength. The photon-number distribution within the bi-stability region and the corresponding Wigner function clarify this point better (middle panels of Fig.~\ref{fig:pump-only}). 
At weaker interaction and within the bi-stability region the number state occupation probability $p_n$ is bimodal and the peak intensities move towards higher photon number as the bi-stable phase is traversed ($\Omega_0 = 6.3,6.5$). Similarly, the corresponding Wigner function has two well-separated local maxima in the $XP$-plane around MF fixed points that are depicted as black and white stars in each case. Upon increasing the interaction, both the occupation probability as well as the phase-space distribution show overlaps between the two states which indicates that the tunneling can indeed be activated via fluctuations, as can be seen in Fig.~\ref{fig:pump-only} for $\Omega_0 = 2.2,~ 2.7$ at $V_0 = 1$.

If we increase the pumping rate even further, the system transitions to a unique-solution phase again (A), as indicated by singly-peaked number-state occupation shown with the blue histograms at the bottom of Fig.~\ref{fig:pump-only}. Unlike the low-power case however, the Wigner function has a banana-shaped distribution indicating the large asymmetry between $X,P$ quadratures. More detailed discussions on the DPT and its relation to the In Appendix~\ref{app:QMC} one can find further discussion about switching dynamics using one quantum Monte-Carlo trajectory and prolonged correlation times within the bi-stability region. 

Next, we investigate the quantum properties of the generated photons by calculating the output spectra of the case investigated in Fig.~\ref{fig:pump-only}(b). The results are shown in Fig.~\ref{fig:pump-only-spectra}(a),(b), via Bogoliubov matrix, on the lower and upper MF branches, respectively. In this case, the parametric process leads to the generation of photon pairs which appear as side peaks in the output intensity spectra, shown with the red and blue lines in Fig.~\ref{fig:pump-only-spectra}(a). 
Notice that here we only focus on fluctuation correlation properties of $\hat{b}$, i.e., on the inelastic part of the spectrum. $S_b(\omega)$ is reminiscent of the Mollow triplet fluorescence spectrum of a coherently driven two-level system at high intensities. 

Upon increasing the pumping rate ($\Omega_0$) the effective detuning between the cavity mode and the laser frequency, as well as the side-band spacing decrease. For spacing less than the linewidth $\gamma_0$, the double-peaked spectrum on the lower branch morphs to a single-peaked feature at the laser frequency $\omega_L$, i.e., the solid green line in Fig~\ref{fig:pump-only-spectra}(a). Increasing the pumping rate further on the upper branch, the effective detuning, and the side-band spacing increases again. This transition can be observed in Fig.~\ref{fig:pump-only-spectra}(b) where the single-peaked spectrum at $\Omega_0 = 1.7$ (solid red line) turns into double-peaked spectra at higher pumping rates (green and cyan lines in Fig.~\ref{fig:pump-only-spectra}(b)). Unlike the lower branch, however, the side-band spacing monotonically increases with increasing the pumping rate hence, no further MF bi-stability is observed. The corresponding output $X,P$-quadratures, shown in Fig.~\ref{fig:pump-only-spectra}(c),(d), for the lower and upper MF branches, respectively, indicate that there is always a partially squeezed quadrature (lines dip below standard quantum limit (SQL) = 0.5). Aside from a finite region around the bi-stability threshold, the quadratures mostly satisfy $\braket{\Delta^2 X}\braket{\Delta^2 P} \approx 0.5$ hence, a minimum-uncertainty state as predicted by MF-Bogoliubov.
%
%
In an open systems whose density matrix dynamics are described via a Liouvillian $\mathcal{L}$ as $\dot{\hat{\rho}} = \mathcal{L}\hat{\rho}$. If there is a steady-state $\hat{\rho}_{ss}$, it be the eigenstate of Liouvillian as $\mathcal{L}\hat{\rho}_{ss} = 0$. Therefore, for a stable dynamics the real part of $\mathcal{L}$ spectrum is upper bounded at $\lambda = 0$. 

\begin{figure}[htbp]
\centering
\includegraphics[width=\linewidth]{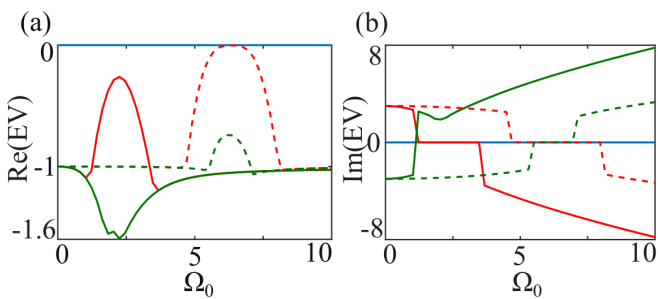}
\caption{~\label{fig:Single-mode Liou.} (a) Real and (b) imaginary parts of the first three eigenvalues (EV) of the Liouvillian as a function of pumping rate $\Omega_0$ in a single mode cavity. The dashed and solid lines correspond to $V_0 = 0.1,1$, respectively at a fixed detuning of $\Delta_0 = -3$. All rates are in terms of $\gamma_0$.}
\end{figure}

To illustrate the closure of the gap at the DPT threshold we calculated the Liouvillian spectrum of the single-mode cavity discussed in the main text. Figure~\ref{fig:Single-mode Liou.}(a),(b) shows the real and imaginary parts of the first three eigenvalues of $\mathcal{L}$ as a function of the pumping rate $\Omega_0$ for $V_0 = 0.1$ (solid lines) and $V_0 = 1$ (dashed lines), respectively. The blue line shows the eigenvalue of the steady-state, i.e., $\lambda = 0$. While far away from MF bi-stability regions the slowest time-scale is set by the cavity decay rate, within the bi-stability range both interactions acquire a slower dynamics, set by an eigenvalue $\Re(\lambda_1) \ge -1$. As can be seen in Fig.~\ref{fig:Single-mode Liou.}(b) within this range the imaginary part of this eigenvalue vanishes as expected from a slowed down dynamics approaching the steady state. 

Interestingly in Fig.~\ref{fig:Single-mode Liou.}(a) the gap between $\lambda_1$ and zero decreases with decreasing the interaction, which implies a frozen dynamics around each MF steady state in the thermodynamic limit. This is also consistent with the tunneling picture and the switching times discussed in Fig.~\ref{fig:pump-only}. 

\section{Quantum Monte-Carlo and the switching rate in first-order DPT}~\label{app:QMC}

As discussed in the main text and described in Appendix~\ref{app:single-mode}, at thermodynamic limit the $1^{st}$-order PT is associated with an abrupt jump at the critical parameter corresponding to multiple MFs. As the system departs from this limit, e.g. by increasing the interaction hence decreasing the particle number, the quantum jumps due to the fluctuations hinder the MF multi-stability picture and lead to a unique solution when the system dynamics is treated fully quantum mechanically. The presence of these local minima however, suggests that the quantum trajectory is mostly probable to be attracted to these fixed points.

\begin{figure}[htbp]
\centering
\includegraphics[width=\linewidth]{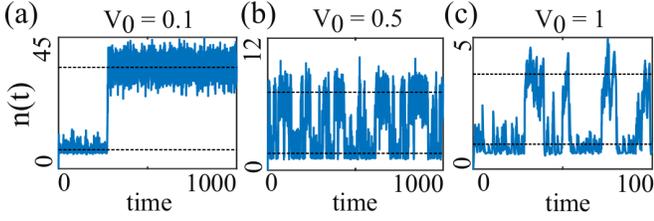}
\caption{~\label{fig:QMC} Particle number as a function of time in a single-mode cavity at different interaction strength and within the MF bi-stability region for (a) $V_0 = 0.1 , \Omega_0 = 6.5$, (b) $V_0 = 0.5 , \Omega_0 = 3$, and (c) $V_0 = 1 , \Omega_0 = 2.2$, determined from one trajectory of a quantum Monte-Carlo simulation. In each panel the black dotted lines show the MF fixed points. The detuning is fixed at $\Delta_0 = -3$. The rates are in units of $\gamma_0$, and the time in units of $\gamma_0^{-1}$.}
\end{figure}

To examine this interpretation further, we employed quantum Monte-Carlo algorithm to investigate a trajectory of a single-mode cavity within its MF bi-stability region. The results are shown in Fig.~\ref{fig:QMC}(a)-(c) for increasing the interaction strength. The black dotted lines in each panel show the two MF solutions while the blue lines are the single quantum trajectory of the system as a function of time. As can be seen the system is switching between these two values with an interaction-dependent rate. While at low interaction (Fig.~\ref{fig:QMC}) the system is barely switching between MF fixed points, the rate noticeably increases upon increasing the interaction. (notice that for $V_0 = 1$ we have shown the zoomed-in dynamics to discern the jumps).

\begin{figure}[htbp]
\centering
\includegraphics[width=\linewidth]{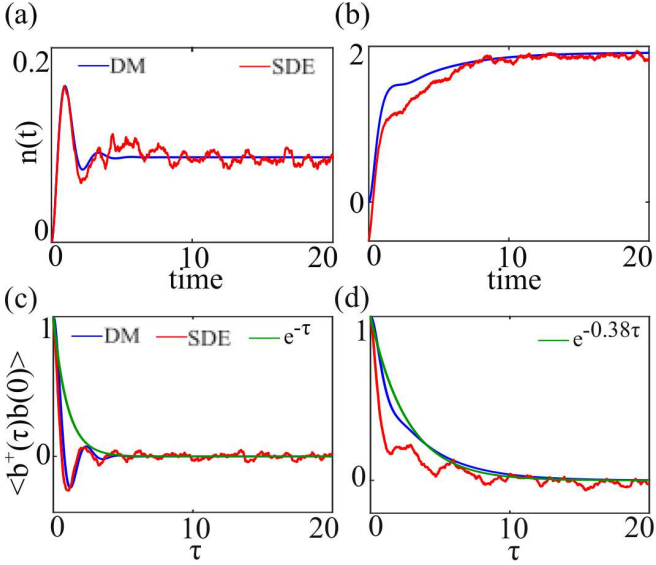}
\caption{\label{fig:SDE vs DM temporal} Instantaneous photon number $n(t)$ in a single-mode cavity as a function of time (in units of $\gamma_0^{-1}$) for $V_0 = 1 , \Delta_0 = -3$ and at (a) $\Omega_0=1$ and (b) $\Omega_0 = 2.2$. (c),(d) Show the first-order correlation of the fluctuations as a function of delay $\tau$. In each panel the solid blue line shows the density matrix calculations (DM) and the red line indicates the stochastic equation (SDE) results. The green lines in panel (c),(d) are the best-squares exponential fits to the correlation function tails.}
\end{figure}

This new time-scale or the tunneling rate can be observed in any temporal dynamics or correlations of observables as well. Figure~\ref{fig:SDE vs DM temporal}(a),(b) shows the photon number relaxation towards the steady-state at strong interaction $V_0 = 1$ for $\Omega_0 = 1, 2.2$, respectively. Figure~\ref{fig:SDE vs DM temporal}(c),(d) shows the behavior of their corresponding first-order correlation $g^{(1)}(\tau)$ as a function of the delay $\tau$. For all cases the dynamics are determined via both full density matrix (DM in solid blue lines) as well as the stochastic differential equations (SDE in solid red lines). As can be seen the results from two approaches agree pretty well and they both predict different time-scales in the unique (Fig.~\ref{fig:SDE vs DM temporal}(a),(c)) and bi-stable (Fig.~\ref{fig:SDE vs DM temporal}(b),(d)) region. While the former shows a fast relaxation towards the steady-state with a rate of $\gamma_0$, the latter has a bi-modal behavior. An exponential fit to the coherence tails, shown in solid green lines in Fig.~\ref{fig:SDE vs DM temporal}(c),(d), indicates that the dynamics starts with a $\gamma_0$-scale behavior. Within the bi-stable region however, the dynamics are slowed down by a factor of 2.5, related to the switching rate between two meta-stable solutions predicated by the MF treatment.

\end{document}